\documentclass[lettersize,journal]{IEEEtran}
\usepackage{amsmath,amsfonts}
\usepackage{algorithmic}
\usepackage{algorithm}
\usepackage{array}
\usepackage[caption=false,font=normalsize,labelfont=sf,textfont=sf]{subfig}
\usepackage{textcomp}
\usepackage{stfloats}
\usepackage{url}
\usepackage{verbatim}
\usepackage{graphicx}
\usepackage{cite}
\usepackage{makecell}
\usepackage{multirow}
\usepackage{xcolor}

\begin{document}

\title{Interlayer Error Calibration for Stacked Intelligent Metasurfaces:\\
Modeling, Algorithms, and Future Perspectives}

\author{Xing Jia, Hao Liu, Haoxian Niu, Jinbao Li, Xiangyu Ding, and Lu Gan
\thanks{This work was supported by the National Natural Science Foundation of China (Grant number 62471096).  (Corresponding author: Lu Gan.) }
\thanks{X. Jia, H. Liu, H. Niu, J. Li, X. Ding, and L. Gan are with the School of Information and Communication Engineering, University of Electronic Science and Technology of China (UESTC), Chengdu, Sichuan 611731. L. Gan is also with the Yibin Institute of UESTC, Yibin, Sichuan 644000, China. H. Liu is also with KU 6G Research Center, Khalifa University, P O Box 127788, Abu Dhabi, UAE (e-mail: xingjia1999@163.com, liu.hao@std.uestc.edu.cn, haoxian\_niu@163.com, li.jinbao@std.uestc.edu.cn, 202422011907@std.uestc.edu.cn, ganlu@uestc.edu.cn).
}}

\maketitle

\begin{abstract}
Stacked intelligent metasurfaces (SIMs) have recently emerged as a key enabler for realizing electromagnetic wave-domain signal processing in next-generation wireless networks.
However, practical SIM implementations often suffer from noticeable mismatches between theoretical models and measured responses due to fabrication and assembly imperfections. This article systematically investigates the problem of interlayer error calibration in SIMs. We first classify representative modeling and hardware-induced imperfections. 
Then, we outline the major challenges in SIM calibration and further develop a general framework that integrates a calibration protocol with the relevant solution strategies.
{
Moreover, we investigate the effectiveness of the multi-stage calibration approach in mitigating geometric deviations and improving the alignment between the calibrated and practical propagation coefficients.}
Finally, we elaborate on key research opportunities and practical challenges toward realizing physically consistent and hardware-compliant SIM implementations for future research.
\end{abstract}

\begin{IEEEkeywords}
Stacked intelligent metasurface (SIM), interlayer calibration, electromagnetic modeling, fabrication imperfections, wave-domain signal processing.
\end{IEEEkeywords}

\section{Introduction}
\IEEEPARstart{T}{he} advent of the sixth-generation (6G) wireless network is expected to transform communications by delivering unprecedented data rates, ultra-low latency, and seamless connectivity for emerging applications such as augmented reality, autonomous driving, and the internet of everything (IoE)~\cite{AnSIM,an2026stacked}. To meet these ambitious requirements, massive multiple-input multiple-output (MIMO) systems are anticipated to further evolve into extremely large antenna array (ELAA) systems. However, this evolution is fundamentally constrained by the heavy reliance of conventional architectures on digital baseband processing and a large number of radio frequency (RF) chains, which results in prohibitive hardware cost, power consumption, and increased processing latency.

{To overcome these challenges, stacked intelligent metasurfaces (SIMs) have emerged as a promising technology, in which multiple metasurface layers are integrated into the base station (BS) to realize wave-domain signal processing with low hardware cost and ultra-fast computational efficiency~\cite{AnSIM,liu2022programmable}.}
Each programmable metasurface layer consists of numerous densely packed subwavelength meta-atoms, which can be dynamically configured via field-programmable gate arrays (FPGAs) to manipulate the incident electromagnetic (EM) waves~\cite{liu2022programmable}. Moreover, SIMs process incident EM signals in the wave domain, thereby enabling parallel analog computations through an EM neural network architecture~\cite{liu2022programmable}. 
{In contrast to conventional BS relying on digital precoding, SIMs can simplify transceiver architectures and reduce the dependency on high-resolution digital-to-analog converters (DACs) and numerous RF chains, which substantially reduces hardware costs and power consumption~\cite{AnSIM,liu2025stacked}. 
}

In wireless communications, An et al. leveraged the wave-domain precoding of SIM to construct multiple decoupled parallel subchannels in the physical space, effectively improving spectral efficiency for holographic MIMO systems~\cite{AnSIM}. 
{
Beyond communication applications, the wave-domain signal processing capability of SIMs has enabled high-resolution sensing and channel estimation. Specifically, Yao et al. developed a channel estimation protocol for SIM-assisted systems, leveraging multiple pilot observations through the SIM to accurately estimate wireless channels~\cite{SIM_Channel}. Moreover, Liu et al. explored SIM-based direction-of-arrival estimation, achieving precise spatial sensing while reducing pilot overhead~\cite{liu2025stacked}.
}
In addition, SIMs can be leveraged as a low-cost analog computing platform for executing complex tasks directly in the wave domain. For instance, Huang et al. utilized an SIM for image classification, achieving over 90\% recognition accuracy without digital baseband processing~\cite{huang2024stacked}. 

Despite these promising advances, most existing studies on SIMs remain at the simulation level, with limited experimental validation. 
{{ A significant challenge for deployment lies in the noticeable mismatches between theoretical model outputs and measured responses, which stem primarily from inevitable fabrication tolerances and assembly imperfections~\cite{liu2022programmable}. First, existing designs typically rely on idealized EM models based on simplified assumptions, which fail to accurately characterize the complex behavior of practical hardware~\cite{AnSIM,11036161,Sun2025Dual-Polarized,Li2025near,10922857}. Second, practical fabrication processes inevitably introduce hardware imperfections, including mechanical misalignment of metasurface layers and defective elements~\cite{Li2025near,Sun2025Dual-Polarized,SIM_Channel}. As EM waves traverse the multiple metasurface layers, these modeling approximations and hardware imperfections accumulate along the stacked structure, causing significant deviations between the ideal and practical interlayer propagation coefficients. This discrepancy fundamentally undermines the accuracy of wave-domain signal processing and leads to severe system performance degradation.}

{While error correction for diffractive deep neural networks has advanced ~\cite{zheng2023dual, chen2023all}, these methods do not directly translate to SIM calibration, since SIMs operate in the microwave band and typically provide only end-to-end measurements with limited internal observability.}
{
This study develops a systematic calibration framework to mitigate errors in the interlayer propagation coefficients for practical SIM deployments.}
We begin by classifying error sources and categories, analyzing both modeling inaccuracies and hardware imperfections. Moreover, calibration protocols and solutions are developed, and the effectiveness of the calibration framework is validated via simulations. Finally, we explore future research directions and summarize the key contributions and prospects.

\section{Interlayer Modeling Errors and Hardware Fabrication Imperfections in SIMs}

 \subsection{Interlayer Modeling Errors}
{Accurate characterization of interlayer EM propagation in an SIM requires a well-defined model that captures the interlayer propagation behavior.
}
However, inconsistencies in assumptions and approximations across different modeling paradigms can cause deviations between ideal and practical responses, leading to inevitable modeling errors. 
{Before elaborating on the interlayer  propagation  model, as shown in Fig.~\ref{fig:modeling}, several key physical parameters of SIM are introduced, which determine the interlayer propagation coefficients.} Existing literature provides several approaches to model the SIM interlayer propagation coefficients, which can generally be categorized into three typical types:

\begin{figure*}[t]
    \centering
    \includegraphics[width=1\linewidth]{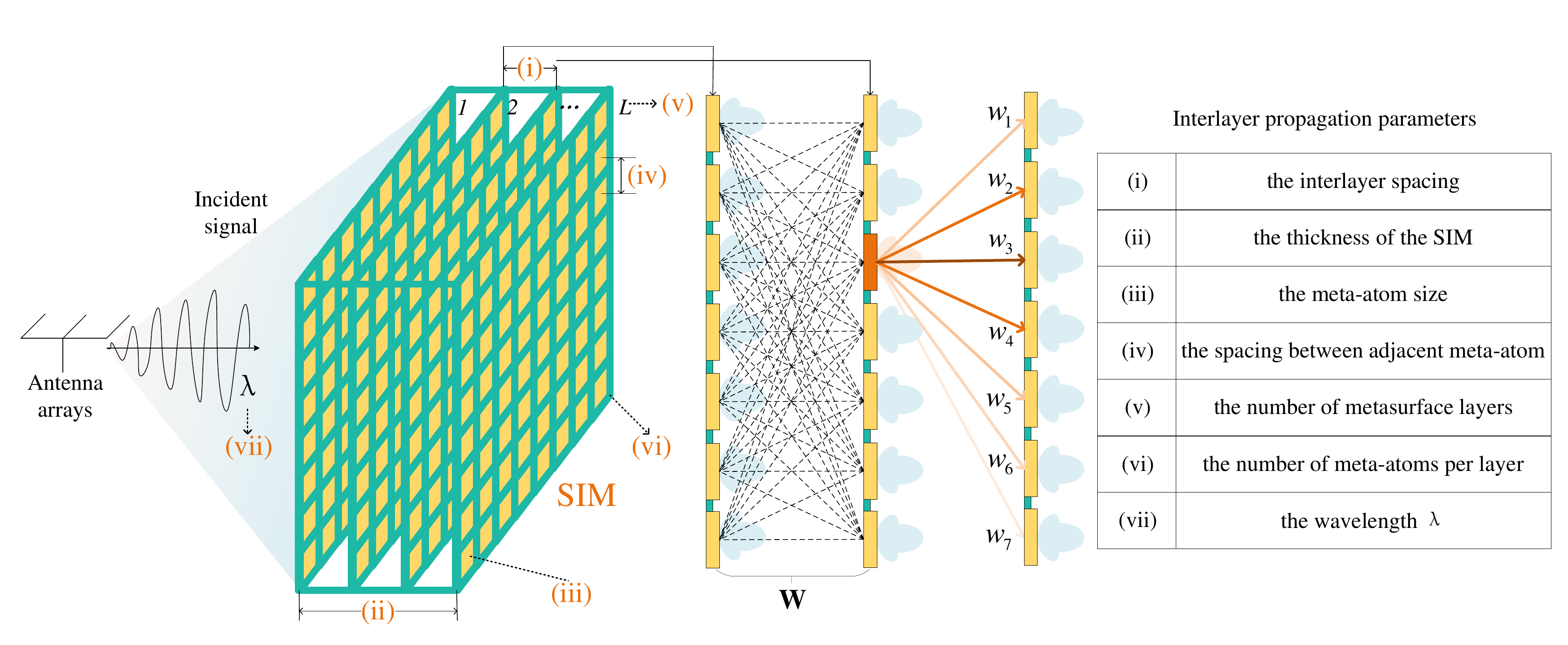}
    \caption{
    Illustration of SIM interlayer signal propagation and the associated key physical parameters (i)–(vii) that govern the interlayer propagation coefficients.
    }
    \label{fig:modeling}
\end{figure*}
 
\begin{itemize}
    \item \textbf{Rayleigh-Sommerfeld Model}: {Such diffraction-based interlayer modeling has been widely adopted in existing SIM studies \cite{AnSIM,Vincenzo2025SIM}. According to the Huygens-Fresnel principle \cite{zhou2021large}, each meta-atom acts as a secondary point source, radiating spherical wavelets that excite every meta-atom on the next metasurface layer.} This model provides an exact solution to the scalar wave equation under point sources and free-space boundary conditions. As a result, it accurately characterizes interlayer propagation in SIMs and serves as the most physically accurate model among existing approaches. However, it exhibits high computational complexity due to the integral's analytical demands, and it becomes inapplicable when the interlayer spacing is extremely compact. Consequently, the model fails to capture mutual coupling effects, leading to significant errors in wavefront manipulation. \textit{Key parameters}: (i), (iii), (iv), and (vii).
    
    \item \textbf{Geometric Radar Model}: This approach models interlayer near-field diffraction by modeling meta-atoms as independent radiators with idealized gain patterns, computing propagation coefficients via a simplified radar equation based on geometric relationships (e.g., distance and antenna gains)\cite{Sun2025Dual-Polarized}. It assumes free-space propagation and neglects mutual coupling and multi-path interference. Unlike the two wave equation-derived models above, this geometric approximation offers computational simplicity but suffers from limited accuracy in dense configurations due to inter-element coupling and boundary effects, leading to phase and amplitude estimation errors. \textit{Key parameters}: (i), (iv), (vi), and (vii).

    \item \textbf{Multiport Network Model}: {In the multiport network model, the SIM is represented as an EM collaborative object, in which each meta-atom on a metasurface layer is modeled as a network port, along with the transmitter and receiver terminals.
    Accordingly, the interlayer propagation behavior is modeled by a global impedance (Z-parameter) matrix with a block structure, which provides a unified representation of the voltages and currents across all ports.
    Compared with conventional modeling, the multiport network model captures interlayer coupling via the impedance-network coupling blocks.
    Consequently, while this model offers an electromagnetically consistent description of interlayer interactions, it may incur prohibitive computational complexity and requires accurate network-parameter identification, especially for large-scale systems. \textit{Key parameters}: (i), (iii), (iv), and (vii).}
\end{itemize}

{
The interlayer propagation models above provide the foundational basis for SIM analysis. However, the idealized approximations in these models inevitably induce modeling errors.
Moreover, practical SIMs also encounter deviations stemming from fabrication imperfections, as detailed next.}

\subsection{Hardware Fabrication Imperfections}

In practice, the performance of SIM often exhibits a moderate gap from the theoretical upper bound predicted by ideal models, primarily due to the inherent challenge of achieving perfect mechanical alignment across multiple layers because of fabrication tolerances, assembly misalignment, and substrate deformation.
Fig.~\ref{fig:hardware_errors} illustrates three representative hardware fabrication errors (\textbf{E}) encountered in practical SIM implementations.

\begin{figure*}[!t]
    \centering
    \includegraphics[width=1\textwidth]{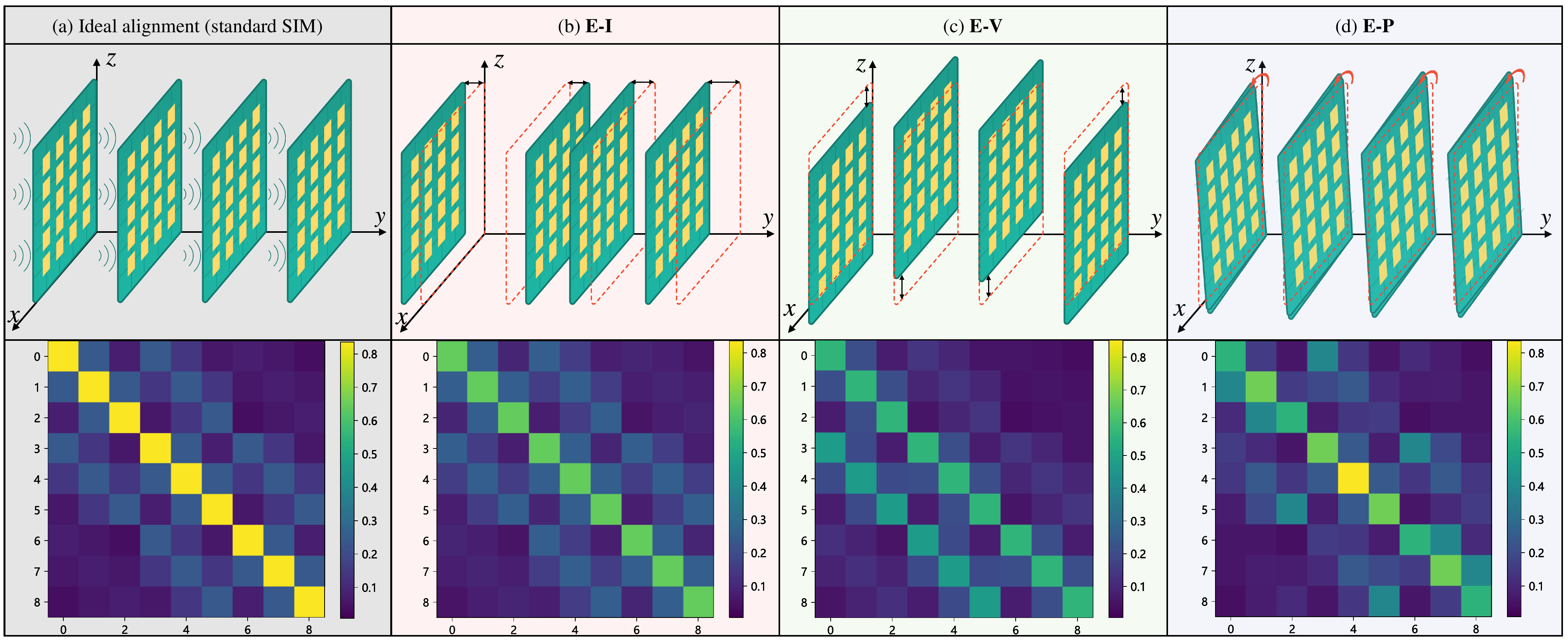}
    \caption{Visualization of three representative hardware fabrication errors in a four-layer SIM architecture and their impact on the induced interlayer propagation coefficient matrices between layers~3 and~4: (a) ideal SIM; (b) {E-I}; (c) {E-V}; (d) {E-P}.}
    \label{fig:hardware_errors}
\end{figure*}

\begin{itemize}

\item \textbf{E-I}: 
Interlayer (\textbf{I})  spacing irregularity refers to variations in the designed separation between adjacent SIM layers along the $y$-axis.
Although uniform spacing is critical for ensuring stable EM response, practical realization is inevitably compromised by bonding tolerances and substrate deformation, which lead to irregular deviations in the interlayer spacing.
As a result, the fixed interlayer spacing becomes irregular, with adjacent layers positioned either closer together or farther apart than intended and thereby altering the designed interlayer  propagation distance.
Furthermore, Fig.~\ref{fig:hardware_errors}(b) shows that {E-I} induces a global increase or decrease in the magnitude of the propagation coefficient matrix entries.

\item \textbf{E-V}: 
Vertical (\textbf{V})  height offset refers to a uniform displacement of an entire metasurface layer along the $z$-axis, which is analogous to the layer being slightly shifted upward or downward relative to its intended plane. 
This phenomenon is generally induced by uneven mounting pressure, substrate warping, or thermal expansion during fabrication. 
As illustrated in Fig.~\ref{fig:hardware_errors}(c), {E-V} introduces a systematic deviation in the interlayer  propagation matrix, producing non-uniform amplitude variations and further disrupting the alignment of the propagating wavefront across adjacent layers.

\item \textbf{E-P}:
{
Planar (\textbf{P}) rotation misalignment arises when a metasurface layer undergoes a slight angular displacement around its geometric center within the metasurface plane. This type of error causes an in-plane rotational offset between adjacent metasurface layers, resulting in angular misalignment among meta-atom arrays that would otherwise be strictly registered across layers.}
It commonly originates from assembly misalignment or twisting of the substrate. 
It is observed from Fig.~\ref{fig:hardware_errors}(d) that {E-P} induces a spatial misregistration and modal distortion within the interlayer propagation coefficient matrix.

\end{itemize}

{In summary, interlayer modeling errors and hardware fabrication imperfections accumulate across multiple SIM layers and are further accompanied by inevitable propagation attenuation, leading to discrepancies between the intended and the practical EM behavior. Such cumulative effects may result in beam-steering inaccuracies, reduced transmission efficiency, and degraded EM manipulation accuracy, thereby revealing an inherent trade-off between enhanced wave-domain processing capability and increased transmission loss as the number of layers grows. To better illustrate these issues, Table~\ref{tab:summary} compares representative SIM works in terms of interlayer  propagation modeling, considered error sources, and optimization variables, showing that most studies evaluate performance under idealized interlayer assumptions, thereby leaving practical SIM calibration errors largely unexplored.
}

\begin{table*}[t]
\centering
\caption{Summary of interlayer  propagation models and meta-atom errors in SIM-assisted systems} \label{tab:summary}
\scriptsize
\setlength{\tabcolsep}{3.5pt}
\renewcommand{\arraystretch}{1.15}
\newcolumntype{C}[1]{>{\arraybackslash}m{#1}}
\begin{tabular}{C{0.5cm} C{1.2cm} C{1.9cm} C{2.4cm} C{1.8cm} C{1.8cm} C{2.0cm} C{2.0cm} C{1.9cm}}
\hline
\textbf{Ref.} & \textbf{Platform} & \textbf{Scenario/Function} & \textbf{Interlayer model} & \textbf{Interlayer error} & \textbf{Meta-atom error} & \textbf{\makecell[l]{Optimization\\variables}} & \textbf{Objective} & \textbf{Method} \\
\hline
\cite{AnSIM} &
\multirow{10}{*}{\makecell[l]{Microwave\\metasurface}} &
MIMO &
Rayleigh-Sommerfeld &
$\times$ &
$\times$ &
\makecell[l]{SIM phases} &
Channel capacity &
Gradient descent \\
\cline{1-1}\cline{3-9}
\cite{liu2022programmable} & &
Image classification &
Measurement-based &
$\times$ &
Quantization &
SIM phases &
Accuracy &
Gradient descent \\
\cline{1-1}\cline{3-9}
\cite{11036161} & &
MIMO &
Multiport network model &
$\times$ &
$\times$ &
SIM phases &
Mean squared error &
Gradient descent \\
\cline{1-1}\cline{3-9}
\cite{Sun2025Dual-Polarized} & &
MU-MIMO &
Geometric radar model &
$\times$ &
$\times$ &
\makecell[l]{SIM phases\\digital precoder} &
Sum-rate &
Alternating optimization \\
\cline{1-1}\cline{3-9}
\cite{Li2025near} & &
MU\mbox{-}MIMO &
Near-field model &
$\times$ &
Phase error &
\makecell[l]{SIM phases\\digital precoder} &
Sum-rate &
Alternating optimization \\
\cline{1-1}\cline{3-9}
\cite{fang2025stacked} & &
MU-MISO &
Rayleigh-Sommerfeld Near-field model &
$\times$ &
$\times$ &
\makecell[l]{SIM phases\\power allocation} &
Max-min rate &
Alternating optimization \\
\hline
\cite{zheng2023dual} &\multirow{5}{*}{\makecell[l]{Optical\\metasurface}} &
Image classification &
Rayleigh-Sommerfeld &
\makecell[l]{{E-I}, {E-P}} &
Phase error &
\makecell[l]{SIM phases\\calibration params} &
Accuracy &
Gradient descent \\
\cline{1-1}\cline{3-9}
\cite{chen2023all} & &
Image classification &
Rayleigh-Sommerfeld &
\makecell[l]{{E-I}, {E-P}} &
Phase error &
SIM phases &
Accuracy &
Gradient descent \\
\cline{1-1}\cline{3-9}
\cite{zhou2021large} & &
Image classification &
Huygens-Fresnel &
\makecell[l]{{E-V}, {E-P}} &
Phase error &
SIM phases &
Accuracy &
Gradient descent \\
\hline
\end{tabular}
\end{table*}

\section{Calibration Protocol and Solution}

\subsection{Calibration Challenges in SIM}
Given these compounded discrepancies, interlayer calibration becomes essential to reconcile the gap between the ideal interlayer diffraction model and practical EM propagation. However, the calibration process faces several key challenges, as summarized below.

\begin{itemize}
    \item \textbf{Hidden interlayer parameters:} Only the end-to-end responses can be directly measured in practice, while the intermediate outputs of each SIM layer remain unobservable. As a result, the interlayer propagation matrices act as hidden variables, leading to an inherent matrix ambiguity problem during optimization.
    
    \item \textbf{High interlayer coupling:} The calibration variables, i.e., the propagation matrices across different layers, are mutually dependent through cascaded EM interactions. Any estimation error in one layer propagates to others, making the optimization landscape highly non-convex and strongly coupled.
    
\end{itemize}

These challenges motivate the development of a structured calibration protocol and effective algorithmic solutions, as presented next.

\subsection{Calibration Protocol}
In the proposed framework, each coherence block is divided into two segments: one dedicated to interlayer calibration and the other to data transmission. Building on this framework, we introduce two mechanisms for invoking calibration, namely periodic calibration and state-driven calibration, as detailed below.

\subsubsection{Periodic Calibration}
Periodic calibration operates according to a fixed time slots configuration, where calibration intervals are inserted at predetermined intervals within the coherence block. This fixed scheduling structure makes the calibration scheme scalable to large deployments and amenable to analytical performance characterization.

Periodic calibration is typically realized in two approaches. The single-stage approach inserts one calibration interval at the beginning of each coherence block, providing a straightforward and low-overhead calibration scheme.
The multi-stage approach, in contrast, divides the coherence block into multiple sequential calibration stages, each followed by a data transmission phase.
In each calibration stage, the test user equipment (UE) transmits pilot symbols that propagate through the SIM toward the BS. 
{To ensure efficient error calibration, random phase configurations are typically employed to modulate the pilot signals across different time slots. This strategy effectively leverages the high degrees of freedom offered by the SIM to generate diverse observation matrices, thereby facilitating accurate propagation coefficient calibration.}
{Specifically, the optimized propagation coefficients from the preceding stage serve as the initial values for the subsequent stage. 
This hierarchical strategy performs coarse global alignment in early stages and fine-tunes local defects in later stages.
Moreover, the framework supports flexible periodicity adjustments to accommodate varying dynamics while maintaining compatibility with single-stage calibration schemes.
}


\subsubsection{State-Driven Calibration}
State-driven calibration is activated only when the system detects notable deviations during the data transmission phase. To this end, the BS continuously monitors several state indicators to assess whether the EM behavior of the SIM remains consistent with its standard operating state.
When the state indicators exceed the predefined threshold, the system switches to a calibration mode.
In this mode, dedicated time slots within the current coherence block are allocated for calibration, during which the test UE transmits pilot signals to recalibrate the SIM and restore its standard operation.
Once the state returns to normal, the system seamlessly resumes data transmission. This protocol enables prompt calibration of emerging discrepancies while maintaining high pilot efficiency.

\subsection{Solution}

\begin{figure*}[t]
    \centering
    \includegraphics[width=\textwidth]{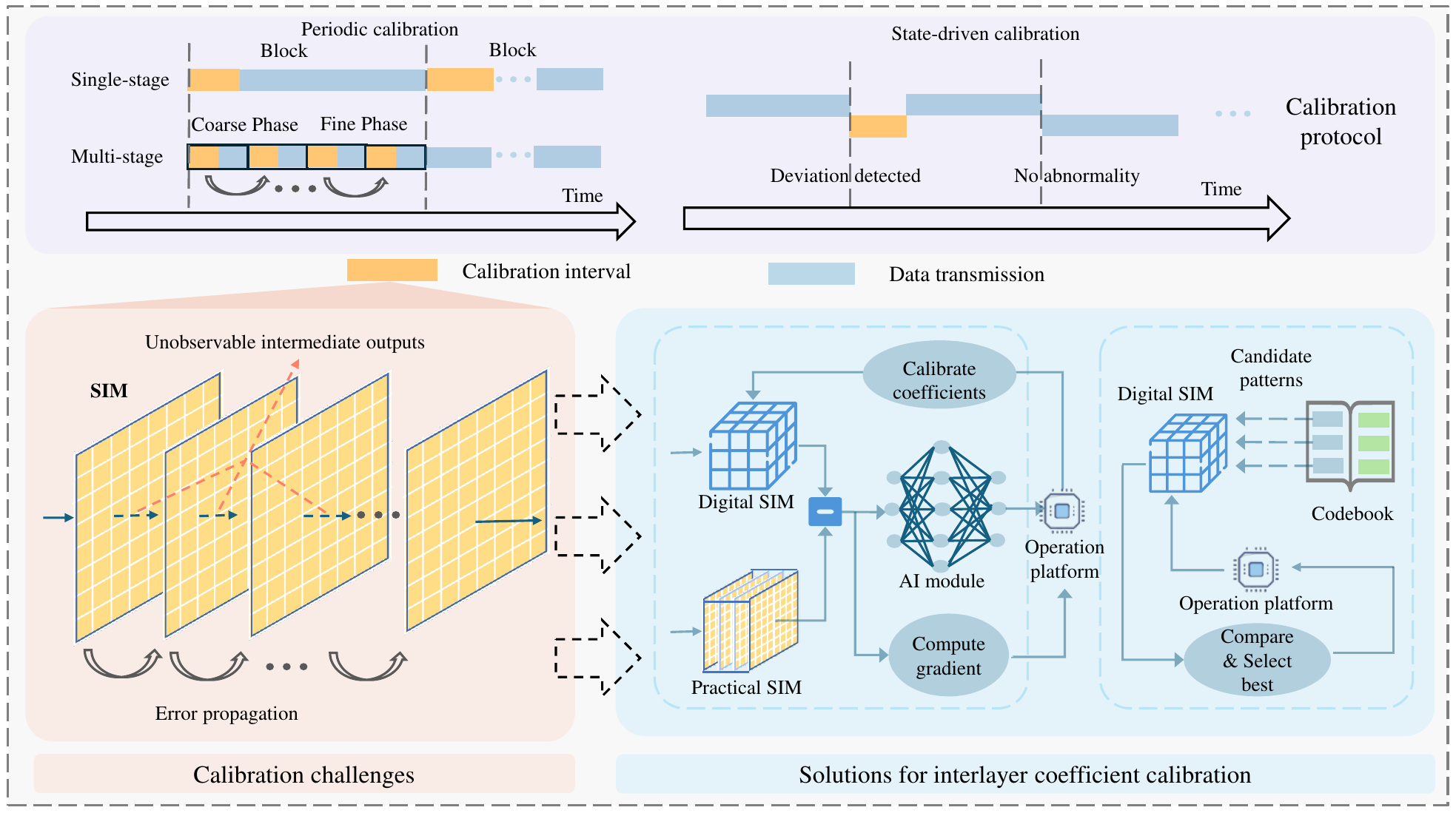}
    \caption{Overview of the SIM calibration framework, including the periodic and state-driven calibration protocols, the main calibration challenges, and the corresponding solution strategies.}
    \label{fig:solutions}
\end{figure*}

\subsubsection{Gradient-Based Calibration Strategy}

This approach compares the practical SIM response with that generated by the ideal model and then computes the gradient of each interlayer coefficient with respect to the normalized mean squared error (NMSE) between them.
The coefficients are updated along the negative gradient direction with a controlled step size, and the process iterates until convergence. 
In practice, the process initializes with the ideal propagation coefficients from the theoretical model. The loop then collects data from pilots, computes the residual, and employs gradient descent to derive a correction value, which is applied to calibrate the theoretical model incrementally. 
The process is compatible with both single-stage and multi-stage calibration  protocols, where coarse stages adopt larger step sizes and simpler patterns, while fine-tuning stages employ smaller steps and more focused configurations. 
This gradient-based framework provides an efficient and scalable approach to keep the SIM’s EM response aligned with its theoretical model, thereby improving operational reliability.

\subsubsection{Codebook-Based Calibration Strategy}
This method relies on an offline codebook that stores a compact set of candidate interlayer propagation coefficients. During each calibration time slot, the system activates a set of candidate patterns, measures their corresponding performance metrics, and selects the index that yields the best response. To enhance search efficiency, the codebook can be designed in a hierarchical structure, where a coarse search first reduces the candidate space before a fine search refines the selection.
This strategy is adaptive, as the codebook allows the system to adjust the search depth according to the allocated calibration duration, thereby achieving a flexible tradeoff between calibration accuracy and overhead.
Moreover, the codebook design offers predictable complexity while providing robustness to noise.

\subsubsection{AI-based Calibration Strategy}
While the gradient-based approach requires complex internal derivative calculations and the codebook strategy often yields compromised performance, the AI-based calibration strategy provides a superior end-to-end solution.
The implementation involves pre-training a neural network offline using practical data and subsequently fine-tuning it online to adapt to the environment.
AI-based strategies are highly advantageous as they bypass the need to compute gradients for internal SIM parameters and avoid considering how to perform backpropagation at the physical layer. Consequently, this paradigm facilitates more powerful inference capabilities and ensures strong robustness against non-ideal hardware characteristics.

\section{Case Studies}
\label{sec:simulation}

\begin{figure}[h]
    \centering
    \subfloat[]{\includegraphics[width=0.48\textwidth]{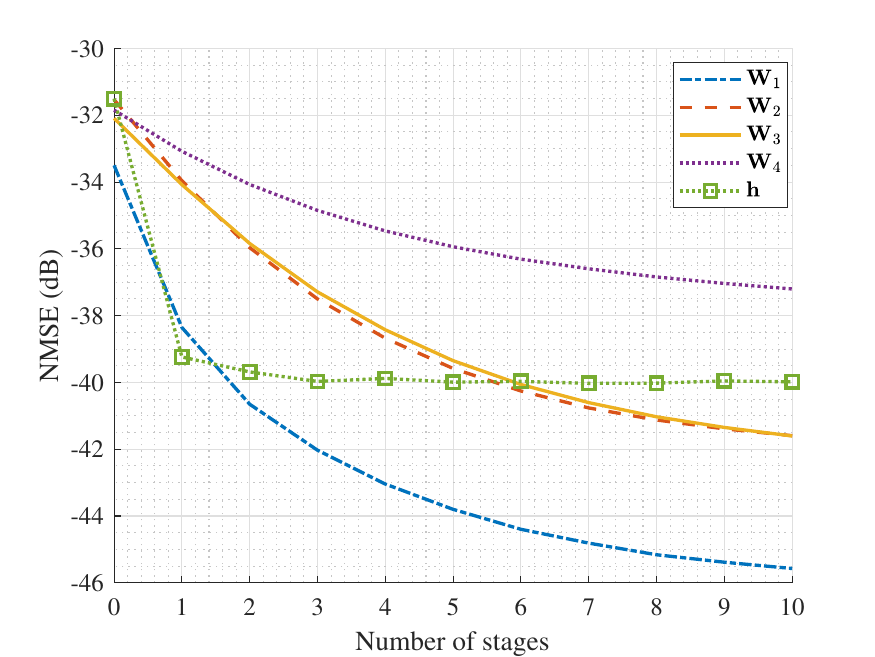}}\\
    \subfloat[]{\includegraphics[width=0.48\textwidth]{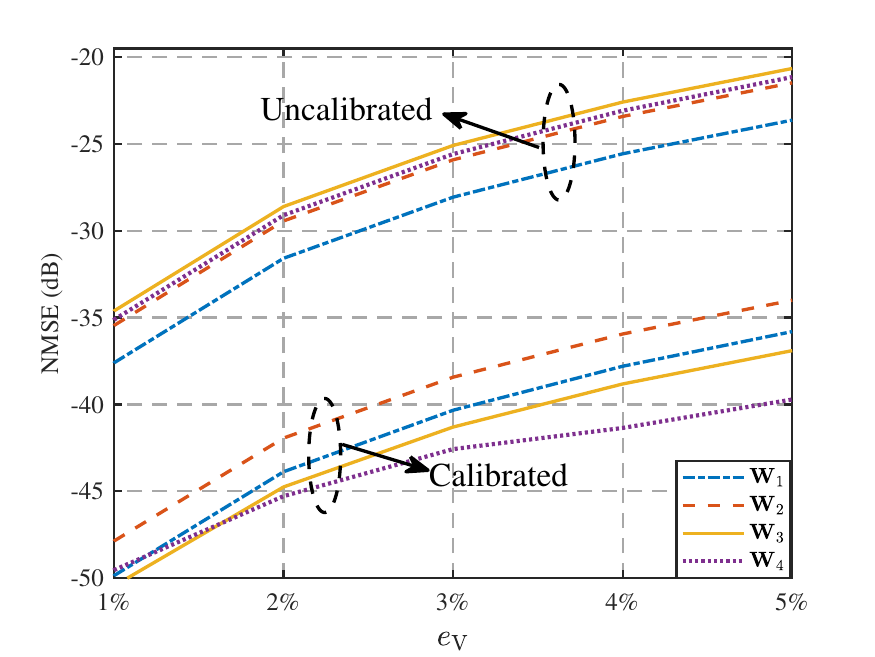}}
    \caption{(a) NMSE between practical and calibrated SIM propagation matrix across stages; (b) NMSE versus the maximum bound of vertical height error $e_\text{V}$ ($6 \times 6$ meta-atom array).}
    \label{fig:sim_vis}
\end{figure}

\begin{figure*}[h]
    \centering
    \includegraphics[width=0.96\textwidth]{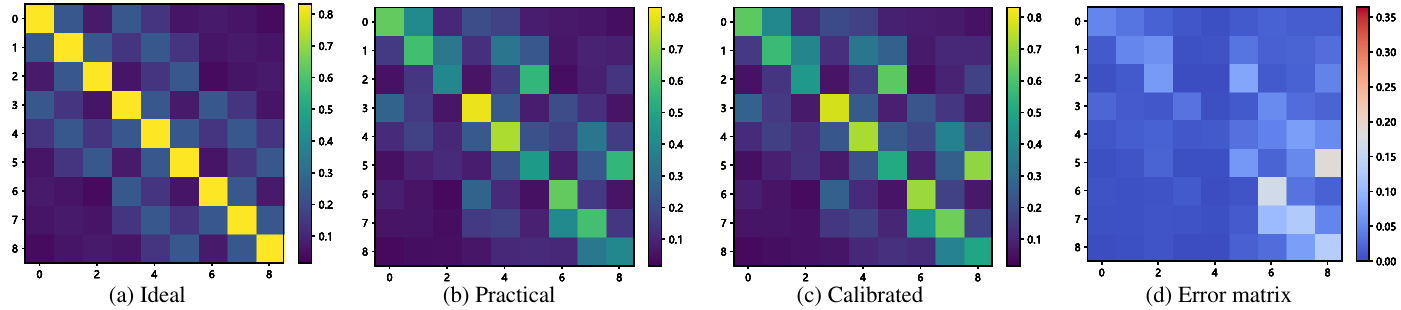} 
    \caption{Visualization of the interlayer propagation coefficient magnitudes from layer $2$ to layer $3$:
(a) ideal magnitude response (Rayleigh–Sommerfeld model);
(b) practical magnitude response;
(c) calibrated magnitude response;
(d) magnitude difference between the practical and calibrated cases ($3 \times 3$ meta-atom array).}
\label{fig:sim_vis_W6} 
\end{figure*}

In this section, we present case studies to evaluate the effectiveness of the calibration framework for SIMs under practical hardware fabrication errors.
{Specifically, we consider a realistic scenario in which a four-layer SIM is integrated with the radome of a BS operating at $28$~GHz, where the wavelength is $\lambda\approx 10.7$~mm, with an overall SIM thickness of $0.01$~m.
For each metasurface layer, the fabrication errors are modeled as independent random variables following uniform distributions within prescribed tolerances: $\text{E-I} \sim \mathcal{U}(-e_\text{I}, e_\text{I})$, $\text{E-V} \sim \mathcal{U}(-e_\text{V}, e_\text{V})$ and $\text{E-P} \sim \mathcal{U}(-e_\text{P}, e_\text{P})$, where the corresponding maximum bounds are $e_\text{I}=0.1\%\,\lambda$, $e_\text{V}=1\%\,\lambda$, and $e_\text{P}=0.01^{\circ}$, respectively.
Moreover, we place a test UE $30$ m away from the BS to transmit pilot signals.
In addition, the calibration protocol follows a periodic multi-stage structure consisting of $10$  calibration stages, with $100$ time slots assigned to each stage.}

{
Fig.~\ref{fig:sim_vis}(a) depicts the calibration evolution of the interlayer  propagation matrices for an SIM composed of uniform $6\times6$ meta-atom arrays. 
Performance is quantified by the NMSE in decibels (dB), which quantifies the relative deviation between the calibrated and practical propagation coefficient matrix, referred to as the calibration NMSE. Consequently, a more negative value indicates more accurate calibration.
Note that the NMSE at the $0$-th stage represents the uncalibrated baseline. 
The distinct initial NMSE levels observed across layers are attributed to the modeling of fabrication imperfections as independent random variables, which results in unique geometric deviations for each layer. 
The progressive refinement across subsequent stages yields a substantial improvement over the first stage (the single-stage benchmark), demonstrating the advantage of multi-stage calibration.
Specifically, at the final stage relative to the $0$-th stage, the NMSE between the practical and calibrated propagation matrices from the first SIM layer to the second is reduced by $12.07$~dB, with an average reduction of $9.25$~dB across all layers, while reducing the UE-to-SIM channel NMSE by $8.48$~dB.
These results demonstrate that the multi-stage calibration framework effectively aligns the calibrated interlayer  propagation matrices with their practical counterparts, thereby compensating  interlayer spacing irregularities, vertical height offsets, and planar rotation misalignment in SIM-based systems.}

{
Fig.~\ref{fig:sim_vis}(b) evaluates the calibration robustness against the vertical height error bound $e_\text{V}$ with $1000$ calibration time slots, where all other error sources are excluded in order to isolate the impact of $e_\text{V}$.
As $e_\text{V}$ increases, the initial NMSE in the uncalibrated system rises significantly. 
In contrast, the calibration algorithm effectively suppresses this error, consistently achieving a significantly lower NMSE floor.
Although the NMSE naturally degrades under larger hardware deviations, this performance loss can be effectively mitigated by increasing the number of calibration time slots, thereby ensuring robustness even under substantial fabrication imperfections.}

{Fig.~\ref{fig:sim_vis_W6} visualizes the interlayer propagation coefficients for a representative SIM layer that exhibits the largest initial NMSE.
For clearer visualization, the metasurface configuration is reset to a compact $3\times 3$ meta-atom array.
Compared with the previous simulation settings, larger fabrication errors are applied.
In Fig.~\ref{fig:sim_vis_W6}(a), the magnitude response of the ideal propagation matrix is computed based on the Rayleigh-Sommerfeld diffraction model. 
Fig.~\ref{fig:sim_vis_W6}(b) presents the practical propagation matrix that incorporates the effects of fabrication and assembly errors, resulting in noticeable deviations from the ideal distribution. 
After applying the proposed calibration algorithm, the calibrated propagation magnitude shown in Fig.~\ref{fig:sim_vis_W6}(c) becomes closely aligned with the ideal case. 
Finally, Fig.~\ref{fig:sim_vis_W6}(d) depicts the magnitude difference between the  practical and calibrated propagation matrix, where the small residual values confirm the effectiveness of the proposed calibration scheme in compensating for hardware-induced distortions. }

\section{Future Research Directions}
\label{sec:future}
Calibration remains a fundamental challenge for practical SIM deployment.
Building upon the proposed framework, several promising research directions can be envisioned to advance the calibration of SIMs.

\subsection{Theoretical Analysis} 
Rigorous theoretical analysis is essential to establish what calibration can deliver under realistic constraints. Progress begins with a clear interlayer model of how one layer influences the next, with assumptions that remain valid when multiple layers operate together. With that foundation, analysis should examine convergence and optimality by investigating whether the update rules converge under realistic noise and pilot patterns, how quickly the objective decreases, and how pilot design and sample size help the procedure avoid poor solutions. It is then natural to quantify performance limits and complexity tradeoffs by first deriving lower bounds on calibration error, using the Cramér–Rao bound (CRB) and Bayesian variants to capture variability and mismatch. These bounds are functions of several parameters, including pilot length, signal-to-noise, and layer/atom counts. Then, they are interpreted alongside computation and memory costs to identify practical operating points.

\subsection{General Errors and Integrated Calibration} 
{In addition to geometric modeling errors, the proposed calibration protocols can be extended to other hardware imperfections. Future research should establish a comprehensive modeling framework that incorporates phase quantization errors and random element impairments, alongside channel estimation inaccuracies and meta-atom pose variation.
Since these uncertainties compound and cascade through the multilayer architecture, isolated calibration approaches are often inadequate, necessitating an integrated calibration framework. Furthermore, developing diagnostic techniques for specific error identification and defective element detection remains a critical open challenge for future investigation.}


\subsection{Measurement of Intermediate States} 
At present, end-to-end calibration schemes are constrained by the difficulty of obtaining accurate interlayer observations in large-scale SIMs. Future research could explore developing hierarchical or partially observable intermediate states that can calibrate propagation characteristics from measured information. The availability of such measurements enables more focused calibration, reduces end-to-end pilot demand, and provides a reliable reference for fast recalibration.

\subsection{Calibration under Multiple Diversity}
{As SIMs evolve toward wideband and multi-polarized operations, the EM response exhibits complex spatiotemporal coupling and nonlinearity that render current narrowband models inadequate. Future research should therefore focus on developing wideband and dual-polarized calibration frameworks that explicitly account for dispersion and cross-orientation mixing. 
To achieve robust generalization, this modeling evolution should be coupled with space-frequency diverse pilot designs and validated against accurate EM models, thereby ensuring accurate calibration while balancing computational complexity.}

\subsection{Sample-Efficient Optimization Methods}
A promising direction is to raise calibration accuracy when only a few pilots are available. The central idea is to represent the system with a compact model that exposes only a small number of parameters. Most of the structure is learned offline and remains fixed, while online updates touch only the parts that are most sensitive to the environment. In practice, this means adding small corrective terms to the calibration model only where needed, so that the model captures the main behavior with far fewer degrees of freedom. Fewer adjustable quantities make each update more stable, thereby reducing the amount of data required to reach a reliable solution and helping to keep computation and memory within real-time limits.

\section{Conclusion}
\label{sec:conclusion}
This article provided a comprehensive overview of interlayer error calibration for SIM, addressing both modeling approximations and practical fabrication imperfections.
Furthermore, we identified the key calibration challenges and presented the calibration protocol along with the solution strategies.
Moreover, simulation results verified that the proposed multi-stage calibration approach effectively aligns the calibrated propagation coefficients with their practical EM counterparts, thereby improving the accuracy of the wave-domain operations performed within the SIM structure.
Finally, we outlined key research directions that connected EM modeling with hardware realization, promoting the advancement of practical and scalable SIM deployments in 6G networks.

\bibliographystyle{IEEEtran}
\bibliography{main}
\end{document}